\documentclass{JHEP3}

\title{Decay Modes of Unstable Strings in Plane-Wave String Field Theory}

\author{Parsa Bonderson\\ California Institute of Technology\\ Pasadena, California 91125\\ E-mail: \email{pbonders@theory.caltech.edu}}

\abstract{The cubic interaction vertex of light-cone string field theory in the plane-wave background has a simple effective form when considering states with only bosonic excitations. This simple effective interaction vertex is used in this paper to calculate the three string interaction matrix elements for states of arbitrary bosonic excitation and these results are used to examine certain decay modes on the mass-shell. It is shown that the matrix elements of one string to two string decays involving only bosonic excitations will vanish to all orders in $\frac{1}{\mu}$ on the mass-shell when the number of excitations on the initial string is less than or equal to two, but in general will not vanish when the number of excitations is greater than two. Also, a truncated calculation of the mass-shell matrix elements for one string to three string decays of two excitation states is performed and suggests that these matrix elements do not vanish on the mass-shell. There is, however, a quantitative discrepancy between this last result and its (also non-vanishing) gauge theory prediction from the BMN correspondence.}

\preprint{hep-th/0307033\\ CALT-68-2444}

\begin{document}

\section{Introduction}

\ 

The recent discovery of a new maximally supersymmetric solution of type IIB
supergravity \cite{Blau:2001ne} has led to a series of interesting
developments. Obtained by taking the Penrose limit of $AdS_{5}\times S^{5}$,
this new solution is a gravitational plane-wave with Ramond-Ramond flux
given by%
\begin{equation}
ds^{2}=-4dx^{+}dx^{-}-\mu ^{2}x_{I}x^{I}\left( dx^{+}\right)
^{2}+dx_{I}dx^{I},\qquad F_{+1234}=F_{+5678}\propto \mu
\end{equation}%
where $I=1,...,8$. Using the light-cone GS formalism, type IIB string theory
in this background was found to be free and exactly solvable \cite%
{Metsaev:2001bj,Metsaev:2002re} despite the non-zero Ramond-Ramond flux
which makes IIB string theory in $AdS_{5}\times S^{5}$ intractable. Applying the Penrose limit allowed for the subsequent extension of the AdS/CFT correspondence to the plane-wave background (which has become known as the BMN correspondence) by the authors of \cite{Berenstein:2002jq}. The significance of the BMN correspondence is largely due to the fact that the exact solvability enables both sides of the correspondence to be studied more explicitly than the AdS/CFT correspondence. In order to incorporate string interactions in the plane-wave background theory, the natural setting to be considered is a generalization of the Minkowski space type IIB light-cone gauge superstring
field theory \cite{Green:1983hw}. The development of the string field theory
cubic interaction vertex for the plane-wave background was carried out in a
series of papers \cite%
{Spradlin:2002ar,Spradlin:2002rv,Schwarz:2002bc,Pankiewicz:2002gs,Pankiewicz:2002tg,He:2002zu}%
, while the extension of the BMN correspondence to the interacting theory
was initiated in \cite%
{Kristjansen:2002bb,Constable:2002hw,Beisert:2002bb,Constable:2002vq}\ and
formulated in \cite{Gross:2002mh,Gomis:2002wi}. In this paper, the string
field theory formalism will be used to calculate the interaction matrix
elements of arbitrary three string states with only bosonic excitations.
These results will be used to consider some properties of string decay modes
on the mass-shell.

\bigskip

\section{Matrix Elements of Bosonic Excitations}

\ 

Restricting to string states $\left| \Psi \right\rangle $ that have only
bosonic excitations, the inner product of these states with the (zero
energy) fermionic vacuum $\left| 0\right\rangle _{b}$\ can be taken
explicitly to leave effective bosonic states $_{b}\left\langle 0|\Psi
\right\rangle =\left| \Psi \right\rangle _{a}$. When reducing to effective
states in this manner, the cubic interaction vertex $\left|
H_{3}\right\rangle $ reduces to an effective bosonic interaction vertex
having the greatly simplified form \cite{Pearson:2002zs}\footnote{%
The BMN basis $\alpha _{\left( r\right) n}^{I}$ is exclusively used
throughout this paper. The relationship between the BMN basis and the
standard oscillator basis is given by $\alpha _{\left( r\right)
0}^{I}=a_{\left( r\right) 0}^{I}$ and $\alpha _{\left( r\right) n}^{I}=\frac{%
1}{\sqrt{2}}\left( a_{\left( r\right) \left| n\right| }^{I}-ie\left(
n\right) a_{\left( r\right) -\left| n\right| }^{I}\right) $ for $n\neq 0$,
where for any $x\in \mathbf{R}$, $e\left( x\right) =\left\{ 
\begin{array}{cc}
1 & \mbox{for}\quad x\geq 0 \\ 
-1 & \mbox{for}\quad x<0%
\end{array}%
\right. $. Also, the Regge slope will be set to $\alpha ^{\prime }=2$ in
this paper.}%
\begin{equation}
\left| H_{3}\right\rangle _{a}=v\mathcal{P}E_{a}\left| 0\right\rangle _{a}
\end{equation}%
\begin{equation}
E_{a}=\exp \left( \frac{1}{2}\sum_{r,s=1}^{3}\sum_{m,n=-\infty }^{\infty
}\sum_{I=1}^{8}\alpha _{\left( r\right) m}^{I\dagger }\widetilde{N}%
_{m,n}^{\left( r,s\right) }\alpha _{\left( s\right) n}^{I\dagger }\right) 
\end{equation}%
\begin{equation}
\mathcal{P}=\frac{\alpha }{2}\sum_{r=1}^{3}\sum_{n=-\infty }^{\infty }\frac{%
\omega _{\left( r\right) n}}{\alpha _{\left( r\right) }}\alpha _{\left(
r\right) n}^{I\dagger }\alpha _{\left( r\right) -n}^{J}\Pi _{IJ}
\end{equation}%
\begin{equation}
\widetilde{N}_{m,n}^{\left( r,s\right) }=\left\{ 
\begin{array}{cc}
\overline{N}_{0,0}^{\left( r,s\right) } & \mbox{for}\qquad m=n=0 \\ 
\frac{1}{\sqrt{2}}\overline{N}_{\left| m\right| ,0}^{\left( r,s\right) } & %
\mbox{for}\qquad m\neq 0,n=0 \\ 
\frac{1}{2}\left( \overline{N}_{\left| m\right| ,\left| n\right| }^{\left(
r,s\right) }-e\left( mn\right) \overline{N}_{-\left| m\right| ,-\left|
n\right| }^{\left( r,s\right) }\right)  & \mbox{for}\qquad m,n\neq 0%
\end{array}%
\right. 
\end{equation}%
\begin{equation}
\alpha _{\left( r\right) }=\alpha ^{\prime }p_{\left( r\right)
}^{+}=2p_{\left( r\right) }^{+}
\end{equation}%
\begin{equation}
\alpha =\alpha _{\left( 1\right) }\alpha _{\left( 2\right) }\alpha _{\left(
3\right) }
\end{equation}%
\begin{equation}
\omega _{\left( r\right) m}=\sqrt{m^{2}+\mu ^{2}\alpha _{\left( r\right)
}^{2}}
\end{equation}%
\begin{equation}
\Pi _{IJ}=\left[ diag\left( \mathbf{1}_{4},-\mathbf{1}_{4}\right) \right]
_{IJ}
\end{equation}%
where it is understood that light-cone momentum conservation in the form $%
\sum\limits_{s=1}^{3}\alpha _{\left( s\right) }=0$ is imposed on the
interaction vertex and that the cubic interaction here enters the full
Hamiltonian with the effective string coupling $g_{2}=4\pi \mu ^{2}\alpha
_{3}^{2}g_{s}$. The overall factor $v$\ is a function of $\mu $, $\alpha
_{\left( 1\right) }$, $\alpha _{\left( 2\right) }$, and $\alpha _{\left(
3\right) }$ which is currently only known (by comparison with gauge theory)
to leading order for large $\mu $ to be $v\left( \mu ,y\right) \simeq 1$.
The Neumann coefficients $\overline{N}_{m,n}^{\left( rs\right) }$ were found
explicitly in \cite{He:2002zu} and their $\frac{1}{\mu }$\ expansions are
provided in the Appendix. The three string bosonic vacuum $\left|
0\right\rangle _{a}=\left| 0_{\left( 1\right) }\right\rangle _{a}\left|
0_{\left( 2\right) }\right\rangle _{a}\left| 0_{\left( 3\right)
}\right\rangle _{a}$ is defined such that $\alpha _{\left( r\right)
n}^{I}\left| 0_{\left( r\right) }\right\rangle _{a}=0$ for all $n$. Notice
that the definition of $\widetilde{N}_{m,n}^{\left( r,s\right) }$ and the
symmetry $\overline{N}_{m,n}^{\left( r,s\right) }=\overline{N}_{n,m}^{\left(
s,r\right) }$ together imply that $\widetilde{N}_{m,n}^{\left( r,s\right) }=%
\widetilde{N}_{-m,-n}^{\left( r,s\right) }=\widetilde{N}_{n,m}^{\left(
s,r\right) }$ for any $m,n,r,s$.

Such a simplified interaction vertex makes it relatively easy to compute
interactions between arbitrary states of only bosonic excitations. It is
clear from the form of the interaction vertex that there must be an even
number of total excitations in order to have a non-zero matrix element.
Consider the three string state $\left| A\right\rangle $ and corresponding
effective bosonic state $\left| A\right\rangle _{a}$ with $2k$ total
excitations (where $k$ is any positive integer) given by%
\begin{equation}
\left| A\right\rangle =A\left| vac\right\rangle
\end{equation}%
\begin{equation}
\left| A\right\rangle _{a}=A\left| 0\right\rangle _{a}
\end{equation}%
\begin{equation}
A=\prod\limits_{j=1}^{2k}\alpha _{\left( r_{j}\right) m_{j}}^{I_{j}\dagger }
\end{equation}%
where $\left| vac\right\rangle =\left| 0\right\rangle _{a}\left|
0\right\rangle _{b}$ is the complete (bosonic and fermionic) three string
vacuum. The states must also satisfy the condition that a physical state $%
\left| \Phi _{\left( r\right) }\right\rangle $ on the $r^{th}$ string
satisfies, which is \cite{Berenstein:2002jq,Spradlin:2002ar}%
\begin{equation}
\sum\limits_{m=-\infty }^{\infty }mN_{\left( r\right) m}\left| \Phi _{\left(
r\right) }\right\rangle =0
\end{equation}%
where $N_{\left( r\right) m}$ is the occupation number operator of the $%
m^{th}$ BMN mode on the $r^{th}$ string (not to be confused with the Neumann
vectors $\overline{N}_{m}^{\left( r\right) }$) given by%
\begin{equation}
N_{\left( r\right) m}=\sum\limits_{I=1}^{8}\alpha _{\left( r\right)
m}^{I\dagger }\alpha _{\left( r\right) m}^{I}+\sum\limits_{a=1}^{8}\beta
_{\left( r\right) m}^{a\dagger }\beta _{\left( r\right) m}^{a}
\end{equation}%
where $\beta _{\left( r\right) m}^{a\dagger }$ and $\beta _{\left( r\right)
m}^{a}$ are the fermionic creation and annihilation operators in the BMN
basis. (Of course, the fermionic terms can be ignored in this paper since
only bosonic excitations are being considered.) Thus, for $\left|
A\right\rangle $ to be a physical state the condition: $\sum%
\limits_{j=1}^{2k}m_{j}\delta _{s,r_{j}}=0$\ must be imposed upon it for $%
s=1,2,3$.

The interaction matrix element of the three string state $\left|
A\right\rangle $ is%
\begin{equation}
\left\langle A|H_{3}\right\rangle ={}_{a}\left\langle A|H_{3}\right\rangle
_{a}=\frac{v\alpha \mu }{2}\frac{1}{\left( 2k\right) !!}\sum\limits_{\sigma
\in S_{2k}}\sum\limits_{j=1}^{k}\mathcal{N}_{\sigma ,j}
\end{equation}%
\begin{equation}
\mathcal{N}_{\sigma ,j}=\left( \frac{\omega _{\left( r_{\sigma \left(
2j-1\right) }\right) m_{\sigma \left( 2j-1\right) }}}{\mu \alpha _{\left(
r_{\sigma \left( 2j-1\right) }\right) }}+\frac{\omega _{\left( r_{\sigma
\left( 2j\right) }\right) m_{\sigma \left( 2j\right) }}}{\mu \alpha _{\left(
r_{\sigma \left( 2j\right) }\right) }}\right) \frac{\widetilde{N}%
_{-m_{\sigma \left( 2j-1\right) },m_{\sigma \left( 2j\right) }}^{\left(
r_{\sigma \left( 2j-1\right) },r_{\sigma \left( 2j\right) }\right) }}{%
\widetilde{N}_{m_{\sigma \left( 2j-1\right) },m_{\sigma \left( 2j\right)
}}^{\left( r_{\sigma \left( 2j-1\right) },r_{\sigma \left( 2j\right)
}\right) }}\Pi ^{I_{_{\sigma \left( 2j-1\right) }},I_{_{\sigma \left(
2j\right) }}}\mathcal{N}_{\sigma }
\end{equation}%
\begin{equation}
\mathcal{N}_{\sigma }=\prod\limits_{j=1}^{k}\widetilde{N}_{m_{\sigma \left(
2j-1\right) },m_{\sigma \left( 2j\right) }}^{\left( r_{\sigma \left(
2j-1\right) },r_{\sigma \left( 2j\right) }\right) }\delta ^{I_{\sigma \left(
2j-1\right) },I_{\sigma \left( 2j\right) }}
\end{equation}%
The overall factor $\frac{1}{\left( 2k\right) !!}$ compensates for the fact
that summing over the entire permutation group $S_{2k}$ overcounts the
distinct terms. Alternatively, $\frac{1}{\left( 2k\right) !!}%
\sum\limits_{\sigma \in S_{2k}}$ in this expression may be replaced by $%
\sum\limits_{\sigma \in G}$ given some $G\subset S_{2k}$ chosen such that $%
\left| G\right| =\left( 2k-1\right) !!$ and $\left\{ \left\{ \sigma \left(
1\right) ,\sigma \left( 2\right) \right\} ...\left\{ \sigma \left(
2k-1\right) ,\sigma \left( 2k\right) \right\} \right\} $ is a distinct
combination of $2$-subsets of the index set $\left\{ 1,...,2k\right\} $ for
each $\sigma \in G$. When restricting this result to excitations along a $%
\mathbf{R}^{4}$ subspace of the plane-wave's transverse $\mathbf{R}^{8}$
directions, one can read off the Feynman rules obtained in \cite%
{Gomis:2003kj}.

\bigskip

\section{On-shell Interactions}

\ 

A peculiar property of this theory is that the three string bosonic
interactions of physical states with less that six total excitations are
either zero, kinematically prohibited, or kinematically suppressed. The term
''kinematically suppressed'' in this context is used to mean that when the
mass-shell condition is imposed on a particular interaction, the $\frac{1}{%
\mu }$ expansion of its matrix element vanishes to all orders. The free
light-cone Hamiltonian operator of the $r^{th}$ string is%
\begin{equation}
H_{2\left( r\right) }=\frac{1}{\alpha _{\left( r\right) }}%
\sum\limits_{m=-\infty }^{\infty }\omega _{\left( r\right) m}N_{\left(
r\right) m}
\end{equation}%
so the mass-shell condition $p^{-}=H_{2}$ combined with conservation of $%
p^{-}$ for a three string interaction of the three string state $\left| \Psi
\right\rangle $ can be expressed as%
\begin{equation}
\sum\limits_{r=1}^{3}\sum\limits_{m=-\infty }^{\infty }\frac{\omega _{\left(
r\right) m}}{\mu \alpha _{\left( r\right) }}N_{\left( r\right) m}\left| \Psi
\right\rangle =0
\end{equation}

For a clear example demonstrating kinematic suppression, consider the cubic
interaction of the three string state $\left| \Psi _{4}\right\rangle =\alpha
_{\left( 1\right) m}^{i\dagger }\alpha _{\left( 1\right) -m}^{j\dagger
}\alpha _{\left( 3\right) n}^{i\dagger }\alpha _{\left( 3\right)
-n}^{j\dagger }\left| vac\right\rangle $ where $i,j$ are distinct with $\Pi
^{ii}=\Pi ^{jj}=1$ and $m,n>0$. The interaction matrix element of this state
is\footnote{%
The notation established in \cite{He:2002zu} is used, where $\approx $
indicates equality to all orders in the $\frac{1}{\mu }$ expansion, i.e.
omitted terms are of order $e^{-2\pi \mu \left| \alpha _{\left( r\right)
}\right| }$. The symbol $\simeq $ is be used to indicate leading order
terms. Also, without loss of generality, a normalization can be taken such
that: $\alpha _{\left( 1\right) }=y$, $\alpha _{\left( 2\right) }=1-y$, and $%
\alpha _{\left( 3\right) }=-1$ where $0<y<1$, and so this will be used where
convenient. When this normalization is used $v$ will be written $v\left( \mu
,y\right) $.}%
\begin{eqnarray}
\left\langle \Psi _{4}|H_{3}\right\rangle  &=&\frac{v\alpha \mu }{4}\left( 
\frac{\omega _{\left( 1\right) m}}{\mu \alpha _{\left( 1\right) }}+\frac{%
\omega _{\left( 3\right) n}}{\mu \alpha _{\left( 3\right) }}\right) \left(
\left( \overline{N}_{\left| m\right| ,\left| n\right| }^{\left( 1,3\right)
}\right) ^{2}-\left( \overline{N}_{-\left| m\right| ,-\left| n\right|
}^{\left( 1,3\right) }\right) ^{2}\right)  \\
&\approx &\frac{v\left( \mu ,y\right) \left( 1-y\right) \sin ^{2}\left( \pi
ny\right) }{2\pi ^{2}\mu \sqrt{\left( 1+\frac{n^{2}}{\mu ^{2}}\right) \left(
1+\frac{m^{2}}{\mu ^{2}y^{2}}\right) }}
\end{eqnarray}%
The mass-shell condition for this interaction is%
\begin{displaymath}
2\sqrt{1+\frac{n^{2}}{\mu ^{2}}}=2\sqrt{1+\frac{m^{2}}{\mu ^{2}y^{2}}}
\end{displaymath}%
\begin{equation}
\Rightarrow y=\frac{m}{n}
\end{equation}%
and imposing the mass-shell condition makes this matrix element vanish to
all orders in the $\frac{1}{\mu }$ expansion. Thus, all of the one string to
two string decay modes are suppressed for a string state with only two or
less bosonic excitations. It is straightforward to show that $0$-mode
excitations on the $3$ string only couple to $0$-mode excitations on the $1$
or $2$ strings and so adding $0$-mode excitations to the above decay process
will also produce decay modes that are not allowed.

In order to demonstrate that all three string interactions are not
suppressed, consider the cubic interaction of the three string state $\left|
\Psi _{6}\right\rangle =\alpha _{\left( 1\right) m}^{i\dagger }\alpha
_{\left( 1\right) -m}^{j\dagger }\alpha _{\left( 2\right) 0}^{k\dagger
}\alpha _{\left( 3\right) n}^{i\dagger }\alpha _{\left( 3\right)
n}^{j\dagger }\alpha _{\left( 3\right) -2n}^{k\dagger }\left|
vac\right\rangle $ where $i,j,k$ are distinct with $\Pi ^{ii}=\Pi ^{jj}=\Pi
^{kk}=1$ and $m,n>0$. The matrix element for this cubic interaction is%
\begin{eqnarray}
\left\langle \Psi _{6}|H_{3}\right\rangle  &=&\frac{v\alpha \mu }{8\sqrt{2}}%
\overline{N}_{0,2n}^{\left( 2,3\right) }\left\{ 2\left( \frac{\omega
_{\left( 1\right) m}}{\mu \alpha _{\left( 1\right) }}+\frac{\omega _{\left(
3\right) n}}{\mu \alpha _{\left( 3\right) }}\right) \left( \left( \overline{N%
}_{m,n}^{\left( 1,3\right) }\right) ^{2}+\left( \overline{N}_{-m,-n}^{\left(
1,3\right) }\right) ^{2}\right) \right.   \nonumber \\
&&\left. \qquad \qquad \qquad \qquad +\left( 1+\frac{\omega _{\left(
3\right) 2n}}{\mu \alpha _{\left( 3\right) }}\right) \left( \left( \overline{%
N}_{m,n}^{\left( 1,3\right) }\right) ^{2}-\left( \overline{N}%
_{-m,-n}^{\left( 1,3\right) }\right) ^{2}\right) \right\}
\end{eqnarray}%
\begin{equation}
\approx \frac{v\alpha _{\left( 2\right) }^{1/2}\left( 3-2\left( \frac{%
\omega _{\left( 1\right) m}\omega _{\left( 3\right) n}}{\mu \alpha _{\left(
1\right) }\mu \left| \alpha _{\left( 3\right) }\right| }\right) -\left( 
\frac{\omega _{\left( 3\right) 2n}}{\mu \left| \alpha _{\left( 3\right)
}\right| }\right) \right) \cos \left( \pi n\frac{\alpha _{\left( 1\right) }}{%
\left| \alpha _{\left( 3\right) }\right| }\right) \sin ^{3}\left( \pi n\frac{%
\alpha _{\left( 1\right) }}{\left| \alpha _{\left( 3\right) }\right| }%
\right) }{2\sqrt{2}\pi ^{3}\mu ^{2}\left| \alpha _{\left( 3\right) }\right|
^{1/2}\left( \frac{\omega _{\left( 1\right) m}\omega _{\left( 3\right) n}}{%
\mu \alpha _{\left( 1\right) }\mu \left| \alpha _{\left( 3\right) }\right| }%
\right) \left( \frac{\omega _{\left( 1\right) m}}{\mu \alpha _{\left(
1\right) }}-\frac{\omega _{\left( 3\right) n}}{\mu \left| \alpha _{\left(
3\right) }\right| }\right) \sqrt{\frac{\omega _{\left( 3\right) 2n}}{\mu
\left| \alpha _{\left( 3\right) }\right| }\left( \frac{\omega _{\left(
3\right) 2n}}{\mu \left| \alpha _{\left( 3\right) }\right| }-1\right) }}
\end{equation}%
To leading order in $\frac{1}{\mu }$ this is%
\begin{equation}
\left\langle \Psi _{6}|H_{3}\right\rangle \simeq -\frac{\left(
m^{2}+3n^{2}y^{2}\right) \left( 1-y\right) ^{1/2}\cos \left( \pi ny\right)
\sin ^{3}\left( \pi ny\right) }{2\pi ^{3}n\left( m^{2}-n^{2}y^{2}\right) \mu 
}
\end{equation}%
Notice that%
\begin{equation}
3-2\left( \frac{\omega _{\left( 1\right) m}\omega _{\left( 3\right) n}}{\mu
\alpha _{\left( 1\right) }\mu \left| \alpha _{\left( 3\right) }\right| }%
\right) -\left( \frac{\omega _{\left( 3\right) 2n}}{\mu \left| \alpha
_{\left( 3\right) }\right| }\right) <0
\end{equation}%
so the only possibility that would allow this matrix element to vanish to
all orders in $\frac{1}{\mu }$\ is if $y=\frac{c}{2n}$ for $c\in \mathbf{Z}%
^{+}$ (i.e. when the overall sine or cosine term vanishes). The mass-shell
condition for this interaction is%
\begin{displaymath}
2\sqrt{1+\frac{n^{2}}{\mu ^{2}}}+\sqrt{1+\frac{4n^{2}}{\mu ^{2}}}=2\sqrt{1+%
\frac{m^{2}}{\mu ^{2}y^{2}}}+1
\end{displaymath}%
\begin{equation}
\Rightarrow y^{2}=\frac{4m^{2}}{\mu ^{2}\left( 2\sqrt{1+\frac{n^{2}}{\mu ^{2}%
}}+\sqrt{1+\frac{4n^{2}}{\mu ^{2}}}+1\right) \left( 2\sqrt{1+\frac{n^{2}}{%
\mu ^{2}}}+\sqrt{1+\frac{4n^{2}}{\mu ^{2}}}-3\right) } \nonumber
\end{equation}%
which to leading order for large $\mu $ is $y\simeq \frac{m}{n\sqrt{3}}$. It
is clear that the condition for kinematic suppression of this interaction is
inconsistent with the mass-shell condition, and so the decay of the state $%
\left| \alpha _{\left( 3\right) n}^{i\dagger }\alpha _{\left( 3\right)
n}^{j\dagger }\alpha _{\left( 3\right) -2n}^{k\dagger }\right\rangle $ into
two strings is allowed. If fact, one can see from the general form of the
three string interactions that the most likely cause of matrix elements
vanishing (to all orders in $\frac{1}{\mu }$) is the sine terms combined
with mass-shell conditions such as $y=\frac{c_{1}}{c_{2}n}$ for $%
c_{1},c_{2}\in \mathbf{Z}^{+}$. Since the mass-shell condition of decay
processes need not satisfy this condition on $y$ in general, this
demonstrates that kinematic supression is not a generic property of the one
string to two string decay modes.

Since the decay of one to two strings is kinematically suppressed for string
states with two excitations, it is of additional interest to determine
whether this is also the case for decays into more than two strings.
Computing the full matrix elements of interactions involving more than three
strings is rather complicated, so instead the truncation scheme of \cite%
{Roiban:2002xr}\ will be employed here with the caveat that the result may
not contain the entirety of the leading order behavior. Specifically, the
truncation that will be used is a restriction of the intermediate string
states to those with two excitations. The process that will be considered
here is the tree level decay from $\left| \alpha _{\left( 4\right)
n}^{i\dagger }\alpha _{\left( 4\right) -n}^{j\dagger }\right\rangle $ to $%
\left| \alpha _{\left( 1\right) m}^{i\dagger }\alpha _{\left( 1\right)
-m}^{j\dagger };0_{\left( 2\right) };0_{\left( 3\right) }\right\rangle $
where $i,j$ are distinct with $\Pi ^{ii}=\Pi ^{jj}=1$, $m,n>0$, and $%
\sum\limits_{r=1}^{4}\alpha _{\left( r\right) }=0$\ (with $\alpha _{\left(
r\right) }>0$ for $r=1,2,3$). This decay process has the mass-shell
condition $\frac{\alpha _{\left( 1\right) }}{\left| \alpha _{\left( 4\right)
}\right| }=\frac{m}{n}$. Letting $\Delta =\frac{1}{E-H_{2}}$, the truncated
iterated $H_{3}$ interation matrix element on the mass-shell is given to all
orders in $\frac{1}{\mu }$ by%
\begin{eqnarray}
&&\left\langle \alpha _{\left( 4\right) n}^{i\dagger }\alpha _{\left(
4\right) -n}^{j\dagger }\right| H_{3}\Delta H_{3}\left| \alpha _{\left(
1\right) m}^{i\dagger }\alpha _{\left( 1\right) -m}^{j\dagger };0_{\left(
2\right) };0_{\left( 3\right) }\right\rangle _{truncated,mass-shell}  \nonumber \\
&&\left. \qquad \approx \left\{ v\left( \mu ,\alpha _{\left( 1\right)
},\alpha _{\left( 1\right) }+\alpha _{\left( 2\right) }\right) v\left( \mu
,\alpha _{\left( 3\right) },\left| \alpha _{\left( 4\right) }\right| \right)
\left( \alpha _{\left( 1\right) }+\alpha _{\left( 2\right) }\right) \right.
\right.   \nonumber \\
&&\left. \qquad \qquad \qquad \left. +v\left( \mu ,\alpha _{\left( 1\right)
},\alpha _{\left( 1\right) }+\alpha _{\left( 3\right) }\right) v\left( \mu
,\alpha _{\left( 2\right) },\left| \alpha _{\left( 4\right) }\right| \right)
\left( \alpha _{\left( 1\right) }+\alpha _{\left( 3\right) }\right) \right\}
\right.   \nonumber \\
&&\left. \qquad \times \frac{\alpha _{\left( 2\right) }\alpha
_{\left( 3\right) }}{16\pi ^{4}\mu ^{2}}\frac{\sin ^{2}\left( \pi n\frac{%
\alpha _{\left( 2\right) }}{\left| \alpha _{\left( 4\right) }\right| }%
\right) }{\left( 1+\frac{n^{2}}{\mu ^{2}\alpha _{\left( 4\right) }^{2}}%
\right) ^{3/2}}\left[ \pi +2\frac{\mu \left| \alpha _{\left( 4\right)
}\right| }{n}~\mathrm{arctanh}\left( \frac{\frac{n}{\mu \left| \alpha
_{\left( 4\right) }\right| }}{\sqrt{1+\frac{n^{2}}{\mu ^{2}\alpha _{\left(
4\right) }^{2}}}}\right) \right] \right. 
\end{eqnarray}%
where $\alpha _{\left( 1\right) }=\frac{m}{n}\left| \alpha _{\left( 4\right)
}\right| $, $\alpha _{\left( 3\right) }=\frac{n-m}{n}\left| \alpha _{\left(
4\right) }\right| -\alpha _{\left( 2\right) }$, and $\alpha _{\left(
2\right) }$ can take all values between $0$ and $\frac{n-m}{n}\left| \alpha
_{\left( 4\right) }\right| $. The following sum evaluation was used in
obtaining the above result\footnote{%
The sum is evaluated by writing $\sin ^{2}\left( \pi p\frac{\alpha _{\left(
1\right) }}{\alpha _{\left( 5\right) }}\right) $\ as $\frac{1}{2}-\left(
-1\right) ^{p}\frac{1}{2}\cos \left( \pi p\left( 1-2\frac{\alpha _{\left(
1\right) }}{\alpha _{\left( 5\right) }}\right) \right) $, applying
appropriate Sommerfeld-Watson transformations to the two resulting sums, and
then using the fact that $F\left[ x\right] =\left( \frac{\coth \left[ \pi
\mu \left( \alpha _{\left( 5\right) }-\alpha _{\left( 1\right) }\right) x%
\right] +\coth \left[ \pi \mu \alpha _{\left( 1\right) }x\right] }{2}\right)
\approx 1$ for $x>0$.}%
\begin{eqnarray}
&&\sum\limits_{p=1}^{\infty }\frac{\sin ^{2}\left( \pi p\frac{\alpha
_{\left( 1\right) }}{\alpha _{\left( 5\right) }}\right) }{\frac{\omega
_{\left( 5\right) p}^{2}}{\mu ^{2}\alpha _{\left( 5\right) }^{2}}\mu
^{2}\left( \frac{\omega _{\left( 4\right) n}}{\mu \left| \alpha _{\left(
4\right) }\right| }-\frac{\omega _{\left( 5\right) p}}{\mu \alpha _{\left(
5\right) }}\right) }\approx \frac{\pi \alpha _{\left( 5\right) }}{4\mu \sqrt{%
1+\frac{n^{2}}{\mu ^{2}\alpha _{\left( 4\right) }^{2}}}}  \nonumber \\
&&\left. \qquad \qquad \qquad \qquad +\frac{\left|
\alpha _{\left( 4\right) }\right| \alpha _{\left( 5\right) }}{2n\sqrt{1+%
\frac{n^{2}}{\mu ^{2}\alpha _{\left( 4\right) }^{2}}}}~\mathrm{arctanh}%
\left( \frac{\frac{n}{\mu \left| \alpha _{\left( 4\right) }\right| }}{\sqrt{%
1+\frac{n^{2}}{\mu ^{2}\alpha _{\left( 4\right) }^{2}}}}\right) \right. 
\nonumber \\
&&\left. \qquad \qquad \qquad \qquad -\frac{\pi \left|
\alpha _{\left( 4\right) }\right| \alpha _{\left( 5\right) }}{2n\sqrt{1+%
\frac{n^{2}}{\mu ^{2}\alpha _{\left( 4\right) }^{2}}}}\frac{\cos \left( \pi n%
\frac{\alpha _{\left( 5\right) }-2\alpha _{\left( 1\right) }}{\left| \alpha
_{\left( 4\right) }\right| }\right) -\cos \left( \pi n\frac{\alpha _{\left(
5\right) }}{\left| \alpha _{\left( 4\right) }\right| }\right) }{\sin \left(
\pi n\frac{\alpha _{\left( 5\right) }}{\left| \alpha _{\left( 4\right)
}\right| }\right) }\right.
\end{eqnarray}%
where $\alpha _{\left( 5\right) }$ represents the intermediate string
momentum. Note that the evaluation of the sum given above is valid
off-shell, and that the term on the second line vanishes upon imposing the
mass-shell condition. Thus, the truncated matrix element is suppressed one
order in $\frac{1}{\mu }$ (compared to the naive leading order of $\frac{1}{%
\mu }$ expected for this term) by the mass-shell condition, and so is not
kinematically suppressed. While this suggests that the decay process does
not exhibit kinematic suppression and so should be an allowed decay mode,
the complete calculation is required for certainty. In fact, it is easy to
see that the naive leading order behavior for intermediate string states
with four bosonic excitations is $\frac{1}{\mu }$ (i.e. this is what it
would be if there are no cancellations that arise in evaluating sums), which
allows for the unlikely possibility of a miraculous cancellation that leads
to kinematic suppression. The main reason for using this particular
truncation scheme is that the restriction to intermediate string states with
two bosonic excitations is thought to correspond to the restriction to
impurity conserving processes in the gauge theory side of the BMN
correspondence. Specifically, if this assumption were correct the leading
order truncated matrix element would be expected to match the corresponding
single-triple trace gauge theory results found in \cite%
{Gursoy:2002fj,Beisert:2002ff,Freedman:2003bh}. Defining $y_{s}=\frac{\alpha
_{\left( s\right) }}{\left| \alpha _{\left( 4\right) }\right| }$\ and
setting $\left| \alpha _{\left( 4\right) }\right| =1$ (without loss of
generality), the leading order term of the matrix element is%
\begin{eqnarray}
&&\left\langle \alpha _{\left( 4\right) n}^{i\dagger }\alpha _{\left(
4\right) -n}^{j\dagger }\right| H_{3}\Delta H_{3}\left| \alpha _{\left(
1\right) m}^{i\dagger }\alpha _{\left( 1\right) -m}^{j\dagger };0_{\left(
2\right) };0_{\left( 3\right) }\right\rangle _{truncated,mass-shell}  \nonumber\\
&&\left. \qquad \qquad \qquad \simeq \frac{\left( \pi +2\right) }{16\pi
^{4}\mu ^{2}}\left( \frac{n+m}{n}\right) y_{2}\left( 1-\frac{m}{n}%
-y_{2}\right) \sin ^{2}\left( \pi ny_{2}\right) \right. 
\end{eqnarray}%
\ Dividing by the state normalization factor $J\sqrt{y_{1}y_{2}y_{3}}$ and
explicitly inserting the factor of $\frac{1}{\mu }g_{2}^{2}$ to make contact
with the gauge theory side of the BMN correspondence produces the string
field theory prediction of the single-triple matrix element on the mass-shell%
\begin{equation}
\left[ \widetilde{\Gamma }_{n;m}^{13}\right] _{SFT}=\frac{g_{2}^{2}}{2\pi
^{2}J\mu ^{2}}\frac{\left( \pi +2\right) }{8\pi ^{2}\mu }\left( \frac{n+m}{m}%
\right) \sqrt{\frac{m}{n}y_{2}\left( 1-\frac{m}{n}-y_{2}\right) }\sin
^{2}\left( \pi ny_{2}\right) 
\end{equation}%
Comparing this result to the gauge theory single-triple matrix element found
in Eqn (5.51) of \cite{Gursoy:2002fj} taken on the mass-shell%
\begin{equation}
\left[ \widetilde{\Gamma }_{n;m}^{13}\right] _{GT}=\frac{g_{2}^{2}}{2\pi
^{2}J\mu ^{2}}\sqrt{\frac{m}{n}y_{2}\left( 1-\frac{m}{n}-y_{2}\right) }\sin
^{2}\left( \pi ny_{2}\right) 
\end{equation}%
does not provide an exact match. This matching fails by a factor of $\frac{%
\left( \pi +2\right) }{8\pi ^{2}\mu }\left( \frac{n+m}{m}\right) $, and
while the $\left( \frac{n+m}{m}\right) $ and numerical factor could
conceivably be compensated for by contributions that were left out by the
truncation, the extra $\frac{1}{\mu }$\ (which can be attributed to the one
order of suppression that was found for the truncated matrix element on the
mass-shell) is problematic. A cursory analysis of the contribution from
intermediate states with four excitations suggests that the complete matrix
element may still contain the appropriate leading order behavior to provide
match the gauge theory results and further confirm the BMN correspondence.
This would imply that the truncation is not appropriate for this process and
brings up the question of when it is valid to apply the truncation scheme.
It should also be mentioned that if the supersymmetry operator $Q_{4}$ is
non-zero, there will be an additional interaction $\frac{1}{2}\left( \left\{
Q_{2},\overline{Q}_{4}\right\} +\left\{ Q_{4},\overline{Q}_{2}\right\}
\right) $ that could potentially contribute to this decay process.

\bigskip

\section{Conclusion}

\ 

The cubic interactions for an arbitrary number of bosonic excitations in the
plane-wave type IIB string field theory computed in this paper provides a
large class of string amplitudes that can be analyzed. The results were used
to compute a $3\rightarrow 2+1$ excitation number decay from one to two
strings and show that it is not kinematically suppressed. It was also
demonstrated that the one to two string decays with bosonic excitations: $%
n_{3}\rightarrow n_{1}+n_{2}$ where $n_{1}+n_{2}=n_{3}>2$ are in general
allowed, whereas none of the one to two string decay modes are allowed for $%
n_{3}\leq 2$. Additionally, the matrix element of the one to three string
decay of two excitation states $2\rightarrow 2+0+0$ was calculated using a
truncation scheme that restricted the intermediate virtual string to states
with two bosonic excitations. This decay was found not to be kinematically
suppressed in this truncation, which suggests that the decay mode is
allowed, however a complete calculation of the matrix element is needed to
be certain. Since the BMN correspondence has already been confirmed to leading order in $\frac{1}{\mu }$ in the three string interaction sector by comparing the pertinent string field theory Feynman diagrams with their corresponding gauge theory Feynman diagrams \cite{Gomis:2003kj}, the remaining interest in verifying the correspondence (to leading order in $\frac{1}{\mu }$) is for interactions of order $g^2_2$ and higher in the string coupling. The failure of the attempt in this paper to verify this correspondence in the one to three string sector using the truncated matrix element provides
further impetus for computing the complete matrix element of this process
and would indicate a deeper relation if the complete matrix element turns
out to confirm the correspondence.

\bigskip

\acknowledgments

\ 

I am very grateful to J. Schwarz and U. G\"{u}rsoy for useful discussions
and comments. The work of P. B. is supported in part by a National Defense
Science and Engineering Graduate Fellowship.

\bigskip

\appendix

\section{Neumann Coefficients}

\ 

For convenience, the $\frac{1}{\mu }$ expansion (to all orders) of the
Neumann coefficients as found in \cite{He:2002zu} will be listed here. Using
the notation%
\begin{equation}
s_{\left( 1\right) m}=s_{\left( 2\right) m}=1,\qquad s_{\left( 3\right)
m}=2\sin \left( \pi m\frac{\alpha _{\left( 1\right) }}{\alpha _{\left(
3\right) }}\right)
\end{equation}%
the Neumann coefficients for $m,n>0$ are%
\begin{equation}
\overline{N}_{m,n}^{\left( r,s\right) }\approx \frac{1}{2\pi }\frac{\left(
-1\right) ^{r\left( m+1\right) +s\left( n+1\right) }}{\alpha _{\left(
s\right) }\omega _{\left( r\right) m}+\alpha _{\left( r\right) }\omega
_{\left( s\right) n}}\sqrt{\frac{\left| \alpha _{\left( r\right) }\alpha
_{\left( s\right) }\right| \left( \omega _{\left( r\right) m}+\mu \alpha
_{\left( r\right) }\right) \left( \omega _{\left( s\right) n}+\mu \alpha
_{\left( s\right) }\right) }{\omega _{\left( r\right) m}\omega _{\left(
s\right) n}}}s_{\left( r\right) m}s_{\left( s\right) n}
\end{equation}%
\begin{equation}
\overline{N}_{-m,-n}^{\left( r,s\right) }\approx -\frac{1}{2\pi }\frac{%
\left( -1\right) ^{r\left( m+1\right) +s\left( n+1\right) }}{\alpha _{\left(
s\right) }\omega _{\left( r\right) m}+\alpha _{\left( r\right) }\omega
_{\left( s\right) n}}\sqrt{\frac{\left| \alpha _{\left( r\right) }\alpha
_{\left( s\right) }\right| \left( \omega _{\left( r\right) m}-\mu \alpha
_{\left( r\right) }\right) \left( \omega _{\left( s\right) n}-\mu \alpha
_{\left( s\right) }\right) }{\omega _{\left( r\right) m}\omega _{\left(
s\right) n}}}s_{\left( r\right) m}s_{\left( s\right) n}
\end{equation}%
for $m>0,n=0$ are%
\begin{equation}
\overline{N}_{m,0}^{\left( r,s\right) }\approx \frac{\left( -1\right)
^{r\left( m+1\right) +s}}{2\pi }\sqrt{\frac{\left| \alpha _{\left( r\right)
}\right| }{\alpha _{\left( s\right) }\omega _{\left( r\right) m}\left(
\omega _{\left( r\right) m}+\mu \alpha _{\left( r\right) }\right) }}%
s_{\left( r\right) m}\qquad \mbox{for}\qquad s\in \left\{ 1,2\right\}
\end{equation}%
and for $m=n=0$ are%
\begin{equation}
\overline{N}_{0,0}^{\left( r,s\right) }\approx \frac{\left( -1\right) ^{r+s}%
}{4\pi \mu \sqrt{\alpha _{\left( r\right) }\alpha _{\left( s\right) }}}%
\qquad \mbox{for}\qquad r,s\in \left\{ 1,2\right\}
\end{equation}%
\begin{equation}
\overline{N}_{0,0}^{\left( r,3\right) }=-\sqrt{\alpha _{\left( r\right) }}%
\qquad \mbox{for}\qquad r\in \left\{ 1,2\right\}
\end{equation}%
All other Neumann coefficients not related to these by the symmetry $\overline{N}%
_{m,n}^{\left( r,s\right) }=\overline{N}_{n,m}^{\left( s,r\right) }$ are
zero.

\end{document}